\documentclass[12pt]{article}
\usepackage{amssymb,amsmath}
\usepackage{graphicx}

\newcommand{\be}{\begin{equation}}
\newcommand{\ee}{\end{equation}}
\newcommand{\bea}{\begin{eqnarray}}
\newcommand{\eea}{\end{eqnarray}}
\newcommand{\beas}{\begin{eqnarray*}}
\newcommand{\eeas}{\end{eqnarray*}}
\newcommand{\ba}{\begin{array}}
\newcommand{\ea}{\end{array}}

\newcommand{\nbox}{{\,\lower0.9pt\vbox{\hrule \hbox{\vrule height 0.2 cm \hskip 0.19 cm \vrule height 0.2 cm}\hrule}\,}}

\def\href#1#2{#2}
\numberwithin{equation}{section}
\newcommand{\Ared}{{\mathcal A}_{\rm red}}

\newcommand{\Sep}{{\mathcal V}}

\begin{document}

\begin{center}

\vspace{1cm} { \LARGE {\bf Separability of Black Holes in String Theory}}

\vspace{1cm}

Cynthia Keeler and Finn Larsen

\vspace{0.8cm}

{\it Department of Physics, University of Michigan,\\
Ann Arbor, MI-48109, USA \\}

\vspace{0.6cm}

{\tt  keelerc, larsenf@umich.edu} \\

\vspace{2cm}

\end{center}

\begin{abstract}
We analyze the origin of separability for rotating black holes in string theory, considering both massless and massive geodesic equations as well as the corresponding wave equations. We construct a conformal Killing-Stackel tensor for a general class of black holes with four independent charges, then identify two-charge configurations where enhancement to an exact Killing-Stackel tensor is possible. We show that further enhancement to a conserved Killing-Yano tensor is possible only for the special case of Kerr-Newman black holes. We construct natural null congruences for all
these black holes and use the results to show that only the Kerr-Newman black holes are algebraically special in the sense of Petrov. Modifying the asymptotic behavior by the subtraction procedure that induces an exact $SL(2)^2$ also preserves only the conformal Killing-Stackel tensor.  Similarly, we find
that a rotating Kaluza-Klein black hole possesses a conformal Killing-Stackel tensor but has no further enhancements.
\end{abstract}

\pagebreak
\setcounter{page}{1}
\reversemarginpar

\section{Introduction}

The study of black holes benefits greatly from the fact that probes in rotating black hole backgrounds often turn out to satisfy separable equations of motion.
For example, the Klein-Gordon equation of a scalar field propagating in the Kerr-Newman black hole background separates into independent radial and polar equations \cite{Kerr:1963ud,Newman:1965my,Carter:1968ks,Carter:1968rr,CarterReview1}. This kind of separability is by no means obvious, since rotating black holes do not possess a sufficient number of conventional symmetries for it to follow automatically. Many years of extensive effort by many researchers has resulted in progress on the formal characterization of separability (e.g. \cite{Carter:1977pq,Stephani:1978,Benenti1979,Stephani:2003tm,Frolov:2008jr,Kubiznak:2009sm,Cariglia:2011uy}) but its physical significance, if any, remains obscure.

Black holes in string theory offer a broader perspective on separability. For starters, these black holes do in fact exhibit separability \cite{Maldacena:1997ih,Cvetic:1997uw,Cvetic:1997xv,Chow:2008fe}. This behavior is even more surprising than the corresponding feature of the Kerr-Newman geometry because the geometry of black holes in string theory appears much more intricate. More importantly, in the string theory setting the microscopic understanding gives an independent approach to symmetries and this could ultimately give insight into separability.

In the present work we consider black holes that are not necessarily near extremality. In this general setting the microscopic structure is not yet well understood, although there are tantalizing clues
\cite{Larsen:1997ge,Castro:2010fd,Cvetic:2011hp,Cvetic:2011dn}. It may be significant that these clues are
tied to separability and its associated symmetries. The subtraction procedure proposed in \cite{Cvetic:2011dn} alters the black hole backgrounds such that their approximate $SL(2)^2$ symmetry in the near horizon region becomes exact. In this context separability is a physical criterion that constrains the subtraction procedure. This circumstance is our main motivation for developing separability further in the context of black holes in string theory.

In four dimensions separability of the massive Hamilton-Jacobi equation, or of the massive Klein-Gordon equation, is equivalent to the existence of four conserved quantities. In the case of radially symmetric spacetimes these are the mass, the energy, the total angular momentum, and the azimuthal angular momentum of the probe. However, rotating black holes are not spherically symmetric so total angular momentum is not conserved. Separability in this case amounts to an additional conservation law not due to the Killing vectors. In the case of Kerr-Newman black holes it has been known for a long time that an additional conserved quantity can be formed from a nontrivial Exact Killing Stackel tensor (EKST) that is symmetric in its two indices. However, this result does not generalize directly to a generic black hole in string theory. Instead we find that in all cases there is a {\it conformal} Killing-Stackel tensor (CKST). A CKST satisfies a weaker condition than an EKST, yet it is sufficient to account for separability of both the {\it massless} Hamilton-Jacobi equation and the {\it massless} Klein-Gordon equation.

The existence of a Killing-Yano tensor (KYT) is an interesting condition that is stronger than the existence of a KST. The KYT is a conserved anti-symmetric two-tensor that squares to a KST. The KYT guarantees full quantum separability of the Klein-Gordon equation and it even implies separability of the Dirac equation \cite{Carter:1977pq,Carter:1979fe}. The Kerr-Newman black hole does in fact have a KYT but we find that more general black holes in string theory do not.
The apparent absence of a KYT for cases more complex than Kerr-Newman is independently interesting in that the KYT is related to worldline supersymmetry (SUSY) of a probe \cite{Gibbons:1993ap,DeJonghe:1996fb}
and one might have hoped that such a weak form of SUSY would survive even for very general black holes in theories with SUSY. It remains possible that some form of generalized KYT can be related to SUSY even in the general case \cite{Kubiznak:2009qi}.

Kerr-Newman spacetimes are special not only in allowing a Killing tensor: their Riemann curvature tensor is of Petrov type D so they are also algebraically special, as required by their possession of a Killing-Yano tensor. The more general black holes we consider are not algebraically special: they are Petrov type I, the generic class in four dimensions. The algebraic class thus proves a good indicator of separability for the Dirac equation but it is too crude to distinguish between weaker kinds of separability.

An illustration is provided by the ``two-charge'' black hole, a subclass of the general ones we consider. These are just Petrov type I and there is no KYT; but there is enhanced symmetry in that the universal CKST can be promoted to an EKST. For these black holes the Klein-Gordon equation indeed has enhanced separability: it can be separated for arbitrary mass rather than just for massless fields.  These black holes continue to be separable at the quantum level.

One way to improve the algebraic classification is to note that the eigenvectors of the Riemann tensor are the in and outgoing null geodesics. These geodesics generate a pair of natural null congruences which in all cases we study are closely related to the universal CKST. The cases where the CKST can be promoted to an EKST
are precisely those where the natural null congruences are shearfree. The existence of pair of shearfree null congruences which diagonalize the radial/temporal CKST is thus a criterion for separability with arbitrary mass.

As we have mentioned, our main motivation for developing the separability of black holes in string theory is that it appears to play a central role in the emergence of near horizon conformal symmetry for general black holes. Our results developed for asymptotically flat black holes do indeed apply also to near horizon geometries with modified asymptotic behavior. In particular, we find that all of the subtracted geometries as defined in \cite{Cvetic:2011hp} exhibit the weakest form of separability, that is the same amount of separability as is present for the generic four-charge black hole.

The four-charge black holes comprising the main examples of this paper are not the most general rotating black holes in $N=4,8$ string theory. Indeed, the most general ones have not yet been constructed explicitly, even up to duality. They can be constructed in principle through the familiar solution generation mechanism and all that is lacking is a manageable parametrization. It is possible that separability is in fact preserved by the solution generation mechanism and so we may inquire whether these general black holes will be separable just like the four-charge black holes we have considered above. As a test of this hypothesis we also find a CKST for the rotating Kaluza-Klein black hole constructed in \cite{Larsen:1999pp} (for a recent review see \cite{Horowitz:2011cq}).

This paper is organized as follows. In section \ref{Summary} immediately below we provide a concise summary of our results. In Section \ref{TheSetting}
we discuss our notation and parameterizations. We then present the details of increasingly restrictive conditions on the metrics.  First, we review the conventional Killing vector structure in Section \ref{Kvectorsection}.  Then we consider CKSTs and EKSTs in Section \ref{KStensorsection}, and shearfree null congruences in Section \ref{shearfreesection}. As the final condition on our metrics, we consider Petrov type and Killing-Yano tensors in Section \ref{KYsection}. Lastly, in Section \ref{magchargesection}, we consider the separability of metrics which add a magnetic KK charge as in \cite{Larsen:1999pp}.

\section{Summary of Results}\label{Summary}
Before turning to the technical aspects of our work it is worth summarizing our results in a concise form.

We consider several levels of separability for black holes, depending on which equations of motion allow separation of variables:
\begin{itemize}
\item
{\it The null geodesic equation} is separable. This is equivalent to separability of the massless Hamilton-Jacobi equation (HJ eqn.) in the background. The examples we consider will satisfy minor additional technical conditions such that the {\it massless Klein-Gordon equation} (KG eqn.) is also separable in these cases.
\item
{\it The timelike geodesic equation} is separable.  This is equivalent to separability of the massive Hamilton-Jacobi equation in the background.
The examples we consider will satisfy minor additional technical conditions such that the {\it massive Klein-Gordon equation} (KG eqn.) is also separable in these cases. In fact, in the cases we consider, the Killing Tensor can be promoted to an operator that commutes with the d'Alembertian, implying {\it quantum separability} of the Klein-Gordon equation.
 \item
 {\it  The Dirac equation} is separable.  Backgrounds which have this separability automatically have separability of the massive Klein-Gordon equation at the quantum level.
\end{itemize}
We associate each of these degrees of separability with a particular Killing object, as indicated in the table below. A background has a checkmark if it possesses the Killing object in question, and dash if it does not. It is also worth repeating that, in all the cases we consider, we can construct a pair of congruences of shearfree null geodesics which diagonalize the radial/temporal CKST if and only if an EKST exists.

\bigskip
\begin{table}[ht]
\centering
\label{resultstable}
\begin{tabular}{|l|ccc|}
\hline
Separability  & massless & massive  & \\
 & HJE + KGE & HJE + KGE  & Dirac eqn.\\
\hline
Killing Object  & CKST & EKST  & KYT\\
\hline
\hline
Kerr-Newman & \checkmark  & \checkmark & \checkmark \\
two-charge & \checkmark & \checkmark  & --- \\
5D mSugra & \checkmark & ---  & --- \\
four-charge & \checkmark & --- & --- \\
\hline
Subtracted & \checkmark & --- & --- \\
\hline
KK Black Hole & \checkmark  & --- & --- \\
\hline
\end{tabular}
\caption{Each row represents a spacetime background.  Columns show whether a given background possesses the indicated Killing object: Conformal Killing-Stackel Tensor (CKST), Exact Killing Stackel Tensor (EKST), or Killing-Yano Tensor (KYT).  In the spacetimes we study, these objects correlate with separability of the Klein-Gordon equation (KGE), Hamilton-Jacobi equation for geodesics (HJE) or Dirac equation as indicated.}
\end{table}
The black hole families referred to in the table as Kerr-Newman, two-charge, 5D mSugra, four-charge, subtracted, and KK Black hole are all introduced in the following section, specifically in the list near the end of section \ref{TheSetting}.

\section{The Setting}\label{TheSetting}
Our main example is the four-charge generating solution of 4D asymptotically flat black holes in $N=4$ string theory \cite{Cvetic:1995kv},
parameterized by mass, angular momentum and four $U(1)$ charges. The physical parameters in turn are parameterized via
\begin{align}
G_4M&=\frac{1}{4}m\sum^3_{I=0}\cosh 2\delta_I~,\notag\\
G_4Q_I&=\frac{1}{4}m \sinh 2\delta_I~,~(I=0,1,2,3)~,\label{parameters}\\
G_4J&=ma\left(\Pi_c-\Pi_s\right)~,\notag
\end{align}
where we use the abbreviations
\begin{equation}
\Pi_c\equiv\prod^3_{I=0}\cosh\delta_I~, \;\; \Pi_s\equiv\prod^3_{I=0}\sinh\delta_I~.\label{abbreviations}
\end{equation}
We write the metric as
\begin{equation}\label{metric}
ds_4^2=-\Delta^{-1/2}G\left(dt + \frac{a\sin^2\theta}{G} \Ared d\phi \right)^2+\Delta^{1/2}\left(\frac{dr^2}{X}+d\theta^2+\frac{X}{G}\sin^2\theta d\phi^2\right)~,
\end{equation}
where we have
\begin{align}
X &= r^2-2mr+a^2~,\notag\\
G &= r^2-2mr+a^2\cos^2\theta~, \label{metricdefs}\\
\Ared  &= 2m\left[\left(\Pi_c-\Pi_s\right)r+2m\Pi_s\right]~,\notag
\end{align}
and
\begin{align}\label{Delta0}
\Delta_0=\prod^3_{I=0} & \left(r+2ms^2_I\right)+2a^2\cos^2\theta\left[r^2+mr\sum^3_{I=0}s^2_I+4m^2\left(\Pi_c-\Pi_s\right)\Pi_s\right.\notag \\
&\left. -2m^2\sum_{I<J<K}s^2_Is^2_Js^2_K\right]+a^4\cos^4\theta~.
\end{align}
We use the notation $s_I^2=\sinh^2\delta_I$.

We will also consider the subtracted versions of these black holes \cite{Cvetic:2011dn} which still have geometry of the form (\ref{metric}) with the identifications (\ref{metricdefs}), but their conformal factor (\ref{Delta0}) is replaced by:
\begin{equation}\label{DeltaS}
\Delta_s=(2m)^3r(\Pi_c^2-\Pi_s^2)+(2m)^4\Pi_s^2-(2m)^2(\Pi_c-\Pi_s)^2a^2\cos^2\theta~.
\end{equation}
We define the effective potential
\begin{equation}\label{Sdef}
\Sep\equiv\frac{\Delta-\Ared^2}{G}~.
\end{equation}
For the asymptotically flat case we have
\begin{equation}
\Sep_0 = r^2+2mr\left(1+\sum^3_{I=0}s^2_I\right)+8m^2(\Pi_c-\Pi_s)\Pi_s-4m^2\sum_{I<J<K}s_I^2s_J^2s_K^2
+a^2\cos^2\theta~, \label{S0}
\end{equation}
while for the subtracted case we find
\begin{equation}
\Sep_s = -4m^2(\Pi_c-\Pi_s)^2~.\label{Ss}
\end{equation}

The subtracted black holes defined by (\ref{DeltaS}) and (\ref{Ss}) exhibit the same thermodynamic behavior as the original ones, but their asymptotic behavior has been modified from flat to a geometry exhibiting an $SL(2)^2$ symmetry by deforming the supporting matter as needed.

Both the original and the subtracted geometries depend on the four charges $Q_I$ with $I=0,1,2,3$. Generically these charges are all distinct and there is no relation between them. In order to find interesting enhancements of symmetry we will consider several special cases of the original geometries:
\begin{itemize}
\item
The {\it two-charge} black holes: the four charges are equal in {\it pairs}, such as $Q_2=Q_3$ and $Q_1=Q_0$.
\item
The {\it 5D msugra} black holes: three of the charges are identical, such as $Q_1=Q_2=Q_3$. These are solutions to the 4D reduction of minimal supergravity in 5D.
\item
The {\it dilute gas} black holes: there is a hierarchy where one charge is much smaller than the others, such as $Q_0\ll Q_{1,2,3}$. This limit gives rise to a decoupled near horizon region. One may focus on the 5D msugra assignment $Q_1=Q_2=Q_3$ without loss of generality.
\item
The {\it Kerr-Newman} black holes: all four charges are identical.
\item
The {\it Kerr} black hole: all four charges vanish.
\end{itemize}

We will gather further evidence that the {\it subtracted geometries} of all of these black holes are generalizations of the dilute gas limit \cite{Cvetic:2011dn}, by showing they possess the same separability structure.

The most general asymptotically flat 4D black hole in $N=4$ string theory is parameterized up to duality by mass, angular momentum and five $U(1)$ charges \cite{Cvetic:1996zq}. Even though it has not yet been constructed explicitly
we can contemplate its separability. As a probe of this question we will consider in Section \ref{magchargesection} the {\it rotating Kaluza-Klein} black holes
\cite{Larsen:1999pp} which are {\it not} a subset of the four-charge black holes, even though they are solutions in the same theory.

\section{Killing Vectors and Their Properties}\label{Kvectorsection}
The stationary axisymmetric spacetime backgrounds we consider in (\ref{metric}) always have two Killing vectors:
\begin{align}
\xi^\mu \partial_\mu &= \partial_t~, \, \, \eta^\mu\partial_\mu=\partial_\phi~,\\
0 &= \nabla_{(\mu}\xi_{\nu)} = \nabla_{(\mu}\eta_{\nu)}~. \label{kveqn}
\end{align}
If $T^\mu$ is the tangent to an affinely-parameterized geodesic, then $T^\mu \xi_\mu$ and $T^\mu \eta_\mu$ are both constants along the path of the geodesic. Additionally the metric provides the constant, $T^\mu T^\nu g_{\mu\nu}$.  However for separability we need one further constant of motion. The main purpose of this paper is to construct such additional constants of motion in the backgrounds (\ref{metric}), for the Hamilton-Jacobi equation, the Klein-Gordon equation, and the Dirac equation.

It is worth noting that the Killing vectors (\ref{kveqn}) commute with each other:
\begin{equation}\label{kvcommute}
[\eta,\xi]^\nu = \eta^\mu \nabla_\mu \xi^\nu-\xi^\mu \nabla_\mu \eta^\nu = 0~.
\end{equation}
In view of the periodicity of the azimuthal angle, the Killing vectors thus generate the Abelian group $\mathbb{R}\times U(1)$.  The Killing vectors of asymptotically flat axisymmetric spacetimes always commute \cite{Stephani:2003tm}, but it is a nontrivial property for spacetimes with other asymptotics. In the present context the stronger property follows because the spacetime can be written in the form (\ref{metric}) with $\Delta$ a function of $r$ and $\theta$. In particular, it also applies to the subtracted backgrounds considered in \cite{Cvetic:2011dn}.

Any spacetime of the form (\ref{metric}) is also {\it orthogonally transitive}. Abstractly, this property refers to the existence of two-spaces orthogonal to the orbits traced out by the Killing vectors of a stationary axisymmetric spacetime. More explicitly, it simply means there are hypersurfaces orthogonal to the Killing vectors $\eta$ and $\xi$, such as those conventionally defined by $r$ and $\theta$. Theorem 19.1 of \cite{Stephani:2003tm} states that a spacetime is orthogonally transitive whenever the Killing vectors satisfy
\begin{equation}\label{orthtrans}
0 = \xi^\mu R_{\mu[\nu}\xi_\rho \eta_{\sigma]} = \eta^\mu R_{\mu[\nu}\xi_\rho \eta_{\sigma]}~.
\end{equation}
This property is indeed satisfied for all backgrounds of the form (\ref{metric}) with arbitrary $\Delta(r,\theta)$. This is unsurprising since orthogonal transitivity is the ability to pick coordinates labelled by the two Killing directions and the two directions orthogonal to them. The block diagonal form of (\ref{metric}) shows that we have already picked such coordinates. Conversely, according to (eqn. 19.17) in \cite{Stephani:2003tm}, a spacetime that is stationary, axisymmetric, and orthogonally transitive will necessarily have such a block diagonal metric when written in so-called Lewis-Papetrou coordinates.  The upshot here is that the subtraction procedure maintains orthogonal transitivity because the metric remains in the form (\ref{metric}).

\section{Killing-Stackel and Conformal Killing-Stackel Tensors}\label{KStensorsection}
It has long been known that the separability of the HJ and KG equations in the Kerr background is due to the existence of a nontrivial Killing tensor satisfying
\begin{equation}\label{KSeqn}
\nabla_{(\mu}K_{\nu\rho)}=0~.
\end{equation}
For $K_{\mu\nu}$ satisfying (\ref{KSeqn}) it follows that $T^\mu T^\nu K_{\mu\nu}$ is preserved along affinely parameterized geodesics. By ``nontrivial'' we mean that $K_{\mu\nu}$ is not simply the metric $g_{\mu\nu}$, nor a symmetrized product of Killing vectors, such as $\eta_{(\mu}\xi_{\nu)}$.  These objects do satisfy (\ref{KSeqn}) but we have already accounted for their contribution to separability, via the constants $T^\mu \xi_\mu$, $T^\mu\eta_\mu$, and $T^\mu T^\nu g_{\mu\nu}$.  These three constants, plus the additional one $T^\mu T^\nu K_{\mu\nu}$, are sufficient to ensure separability of the HJE.

Importantly, the properties of Kerr do {\it not} immediately generalize to the broader class of charged metrics (\ref{metric}). While the Kerr geometry possesses a nontrivial $K_{\mu\nu}$ satisfying (\ref{KSeqn}), generic members of the class we consider will only have a {\it conformal} Killing-Stackel tensor satisfying
\begin{equation}\label{confkilleqn}
\nabla_{(\mu}K_{\nu\rho)}=g_{(\mu\nu}V_{\rho)}~.
\end{equation}
with nonzero $V_\rho$.  For these objects, $T^\mu T^\nu K_{\mu\nu}$ is only conserved along affinely parameterized {\it null} geodesics.  Thus, a conformal Killing-Stackel tensor only guarantees separability of the {\it massless} Hamilton-Jacobi equation.

The analogous distinction holds for the Klein-Gordon equation as well: the existence of a nontrivial conformal Killing-Stackel tensor ensures separability of the massless Klein-Gordon equation, while separability of the massive equation is only guaranteed by an exact Killing-Stackel tensor solving (\ref{KSeqn}). Related to this, we note in advance that in this section the primary logic will be to start from separability and then identify suitable conserved tensors. However, we should note that the reverse result is also valid: the presence of the CKST or EKST implies separability. For these questions we refer to \cite{Benenti1979,Cariglia:2011uy,Demianski1980,Benenti1997}.

It is worth stating that throughout this paper we make no distinction between a Killing Tensor and a Killing Stackel tensor. Some authors do make such distinctions, but there seems to be no consensus on terminology. In our work the main distinction is between the Conformal KST satisfying (\ref{confkilleqn}) and the Exact KST satisfying (\ref{KSeqn}). We will address the issue of quantum separability in Section \ref{quantumsection}.

\subsection{Constructing Conformal Killing-Stackel Tensors}\label{ConstructingSection}
We can construct a family of conformal Killing-Stackel tensors (CKSTs) for the metrics (\ref{metric}) by adapting the procedure of Davis
\cite{Davis:2006cs} from five dimensions to four dimensions.

Any stationary axisymmetric metric exhibiting orthogonal transitivity can be presented in a block diagonal form with $t,\phi$ and $r,\theta$ components, but no crossterms between these blocks.
The $r,\theta$ part can be further represented as
\begin{equation}\label{rthetametric}
ds_{r,\theta}^2=\Omega(r,\theta)\left(\Theta^2(\theta)d\theta^2+R^2(r)dr^2\right)~,
\end{equation}
for some functions $\Omega,\Theta,R$ with the given dependencies. We have indeed chosen such coordinates to express (\ref{metric}).
In addition to this coordinate choice, we also make the essential and quite nontrivial assumption that the inverted metric is conformally related to one which can be written as a sum of $\theta$ and $r$ dependent pieces:
\begin{equation}\label{uppermetricseparable}
\Omega(r,\theta)g^{\mu\nu}=P^{\mu\nu}(\theta)+S^{\mu\nu}(r)~.
\end{equation}
We can immediately see this condition forces the massless HJ equation,
\begin{equation}
g^{\mu\nu}\partial_\mu S \partial_\nu S=0~,
\end{equation}
to separate, as we can freely multiply by $\Omega(r,\theta)$ in the massless case.

Under the assumptions (\ref{rthetametric}) and (\ref{uppermetricseparable}), $P^{\mu\nu}$ and $S^{\mu\nu}$ are both conformal Killing tensors. Thus they satisfy (\ref{confkilleqn}), in this case with nonzero associated vector fields given by $V$ and $U$ respectively:
\begin{equation}\label{VandU}
V^\mu=(\partial_{\theta}\Omega)g^{\theta\theta}\delta^{\mu}_{\theta}~,~ \; \; U^\mu=(\partial_r\Omega)g^{rr}\delta_r^{\mu}~.
\end{equation}

The CKSTs $P^{\mu\nu}$ and $S^{\mu\nu}$ differ by $\Omega g^{\mu\nu}$, by their definition (\ref{uppermetricseparable}); so they are in the same equivalence class. Specifically, they give rise to the same separation constant in the massless
HJ equation.

\subsubsection{Separability of the Klein-Gordon equation}
In order to establish the separability of the Klein-Gordon equation
\begin{equation}\label{KGequation}
(\nabla^2 -m^2 )\Phi=\frac{1}{\sqrt{-g}}\partial_\mu\left(\sqrt{-g}g^{\mu\nu}\partial_\nu \Phi\right)-m^2 \Phi=0~,
\end{equation}
we will have to do a bit more work. We expand this equation as
\begin{equation}\label{KGexpanded}
\partial_\mu\left(\Omega g^{\mu\nu}\partial_\nu \Phi\right)+\left(\partial_\mu \log \left[\Theta R\sqrt{-\tilde{g}}\right]\right)\Omega g^{\mu\nu}\partial_\nu \Phi-m^2\Omega\Phi=0~,
\end{equation}
where we have multiplied by $\Omega$ and used the relation
\begin{equation}
\sqrt{-g}=\Omega \Theta R \sqrt{-\tilde{g}}~
\end{equation}
between the determinant $g$ of the full metric and the determinant $\tilde{g}$
of the $t,\phi$ block of the metric.

For the massless case, we need only concern ourselves with the first two terms in (\ref{KGexpanded}). Since we have assumed that the
first term is separable via (\ref{uppermetricseparable}) the potential obstacle is the second term. Using the coordinate condition (\ref{rthetametric}), we rewrite this term as
\begin{equation}
\left(\partial_\theta \log \left[\Theta R \sqrt{-\tilde{g}}\right]\right)\frac{1}{\Theta^2}\partial_\theta \Phi + \left(\partial_r \log \left[\Theta R \sqrt{-\tilde{g}}\right]\right)\frac{1}{R^2}\partial_r \Phi~.
\end{equation}
The coordinate dependencies indicated in
(\ref{rthetametric}) show that this term is separable if
\begin{equation}\label{KGsepcond}
\tilde{g} = \tilde{g}_R(r)\tilde{g}_\Theta(\theta)~,
\end{equation}
that is, if $\tilde{g}$ is a product of a function of $r$ and a function of $\theta$. In fact one can always choose coordinates for any axisymmetric spacetime with orthogonal transitivity such that (\ref{KGsepcond}) is satisfied, simultaneously with the condition (\ref{rthetametric}) (e.g. Weyl's canonical coordinates, as in eqn. 19.21 of \cite{Stephani:2003tm}).

In summary, we can choose coordinates for all stationary, axisymmetric spacetimes with orthogonal transitivity, such that the nontrivial condition (\ref{uppermetricseparable}) indicates the existence of a CKST, as well as separability of both the massless HJ and KG equations.

\subsubsection{Explicit Construction of CKSTs}
We can make these results explicit in the context of the black holes with metric given in (\ref{metric}). The first coordinate condition (\ref{rthetametric}) is clearly satisfied since the $r,\theta$ part of the metric is
\begin{equation}
ds^2_{r,\theta} = \Delta^{1/2}\left(\frac{dr^2}{X}+d\theta^2\right)~,\label{4chrtheta}
\end{equation}
while the second coordinate condition (\ref{KGsepcond}) is satisfied since our metrics (\ref{metric}) have $\tilde{g}=-X\sin^2\theta$ where $X$ depends on $r$ alone.

The coordinates used in the form (\ref{uppermetricseparable}) for the inverse metric are thus good coordinates in which to explore both the existence of CKSTs and separability of the massless HJ and KG equations.

After some manipulations we are able to express the inverse metric as
\begin{equation}
\Delta^{1/2}g^{\mu\nu} \partial^\mu\partial^\nu= X\partial_r^2+\partial_{\theta}^2+\frac{1}{\sin^2\theta}\partial_{\phi}^2-\Sep\partial_t^2-\frac{1}{X}\left({\Ared}\partial_t+a\partial_\phi\right)^2~, \label{4chinvmetric}
\end{equation}
where we have used $G=X-a^2\sin^2\theta.$ Since $\Ared$ and $X$ depend only on $r$ this almost takes the separable form (\ref{uppermetricseparable}) with $\Omega=\sqrt{\Delta}$. The only additional condition we must impose is that the
effective potential be separable into two terms as
\begin{equation}\label{Ssep}
\Sep=\Sep_\Theta(\theta)+\Sep_R(r)~.
\end{equation}
The metric (\ref{metric}) will then satisfy the condition (\ref{uppermetricseparable}), with the tensors
\begin{align}
P^{\mu\nu}(\theta) \partial_\mu \partial_\nu &= \partial_{\theta}^2+\frac{1}{\sin^2\theta}\partial_{\phi}^2-\Sep_\Theta\partial_t^2+2a\partial_t\partial_{\phi}~,
\label{P4d}\\
S^{\mu\nu}(r)  \partial_\mu \partial_\nu &= X\partial_r^2-\Sep_R\partial_t^2-\frac{1}{X}\left(\Ared\partial_t+ a\partial_\phi\right)^2-2a\partial_t\partial_{\phi}~,
\label{S4d}\end{align}
and the conformal factor
\begin{equation}\label{Omegadef}
\Omega(r,\theta)=\Delta(r,\theta)^{1/2}~.
\end{equation}
Thus $P^{\mu\nu}$ and $S^{\mu\nu}$ are CKSTs with inhomogeneous terms given by
\begin{align}
V^{\mu} &= \frac{\partial_\theta(\Delta^{1/2})}{\Delta^{1/2}}\delta_\theta^\mu~,\label{Veqn}\\
U^{\mu} &= \frac{\partial_r(\Delta^{1/2})X}{\Delta^{1/2}}\delta_r^\mu~.\label{Ueqn}
\end{align}
The two members of the class of nontrivial CKSTs constructed here apply to the general asymptotically flat backgrounds with four distinct charges as well as to the subtracted backgrounds defined by (\ref{DeltaS}), since both $\Sep_s$ and $\Sep_0$ satisfy (\ref{Ssep}).

The division of the inverse metric (\ref{uppermetricseparable}) into two terms is ambiguous: we can add a constant term to any of the five allowed components of $P^{\mu\nu}$ provided we similarly subtract it from $S^{\mu\nu}$. We used this freedom to add the constant term $2a\partial_t\partial_{\phi}$ in (\ref{P4d}) and subtract the identical term in (\ref{S4d}). We chose this particular constant for later convenience, specifically our construction of principal null vectors in (\ref{Sintoln}).

The key step in the construction of these CKSTs is the separation of $\Sep$ into $\Sep_\Theta(\theta)$ and $\Sep_R(r)$ as in (\ref{Ssep}). In contrast, the conformal factor $\Omega=\sqrt{\Delta}$ can be a complicated function of $r$ and $\theta$. This situation is exactly the same separability requirement as for the massless Klein-Gordon equation identified in \cite{Cvetic:2011dn}.

\subsection{Construction of Exact Killing-Stackel Tensors}
A true Killing tensor satisfies (\ref{KSeqn}), the special case of the CKST equation (\ref{confkilleqn}) where the associated vector $V_\mu$ vanishes. We will refer to such ``true'' Killing tensors as Exact Killing Stackel Tensors (EKSTs).

The CKSTs constructed in the previous subsection can be promoted to EKSTs when the conformal factor $\Omega$ is separable \cite{Davis:2006cs}:
\begin{equation}\label{Omegasep}
\Omega(r,\theta)=\Delta(r,\theta)^{1/2}= f(\theta)+h(r)~.
\end{equation}
Explicitly, in this situation
\begin{equation}\label{fullKS}
K^{\mu\nu}=P^{\mu\nu}-f(\theta)g^{\mu\nu} = h(r)g^{\mu\nu} - S^{\mu\nu}
\end{equation}
is an EKST satisfying (\ref{KSeqn}). The two expressions are equivalent to each other and to
\begin{equation}\label{fullKST}
K^{\mu\nu}(r,\theta)= \frac{1}{\Omega(r,\theta)} \left[ h(r) P^{\mu\nu}(\theta) - f(\theta) S^{\mu\nu}(r)\right]~.
\end{equation}
We have indicated the dependencies of each function in the formula to emphasize the intricate mixing of radial and angular variables in the Exact Killing Tensor.

We already remarked on the immediate relation between the existence of a CKST, separability of the conformally-related inverse metric $\Omega g^{\mu\nu}\partial_\mu \partial_\nu$, and separability of the massless Hamilton-Jacobi equation. The massive Hamilton-Jacobi equation presents an additional obstacle because it involves the operator $g^{\mu\nu}\partial_\mu \partial_\nu - m^2$ rather than just $g^{\mu\nu}\partial_\mu \partial_\nu$. Thus the effective mass term $m^2 \Omega$ must be separated, in addition to the terms already present in the massless case.

The separation of this effective mass term requires precisely the condition (\ref{Omegasep}) which implies the existence of an EKST. We can also see from (\ref{KGexpanded}) that the massive Klein-Gordon equation also becomes separable exactly when (\ref{Omegasep}) is satisfied.

In order to make these considerations explicit we consider for now just the unsubtracted (asymptotically flat) black holes for which the $\Delta_0$ given in (\ref{Delta0}) can be recast as
\begin{align}
\Delta_0 =\left(G+\Ared\right)^2 &+2 m r G \left(\sum^3_{I=0}s^2_I-2(\Pi_c-\Pi_s-1)\right)\notag\\&
-4m^2 G \left(\sum_{I<J<K}s_I^2s_J^2s_K^2-2\Pi_s(\Pi_c-\Pi_s-1)\right)~.\label{Delta0rewritten}
\end{align}
This expression shows that generally $\Omega=\Delta^{1/2}_0$ does not factorize as in (\ref{Omegasep}).
However, for some charge assignments there will be factorization, and in those cases we can construct an EKST. In the following we consider a few special cases.

\subsubsection{Kerr-Newman and the Two-Charge Black Holes}
We first consider the two-charge case, where we set $\delta_3=\delta_2$ and $\delta_0=\delta_1$.  This case also includes the Kerr-Newman black holes, where we set all four charges equal.

Once we make the charges equal in pairs the last two terms in (\ref{Delta0rewritten}) vanish and so we have simply
\begin{align}\label{Omega2chred}
\Omega_{\rm 2ch}=\Delta_{0,{\rm 2ch}}^{1/2}=G+\Ared = X(r)+\Ared(r) - a^2\sin^2\theta~.
\end{align}
Thus $\Omega_{\rm 2ch}$ takes the separated form posited in (\ref{Omegasep}). We choose the separation constant such that
\begin{equation}
f(\theta) = a^2\cos^2\theta~,~~ h(r) = X+\Ared-a^2~.
\end{equation}
Thus we can construct an EKST in the two-charge case, by inserting the appropriate expressions into (\ref{fullKST}).

Using (\ref{Omega2chred}) and the definition of the effective potential $\Sep$ in (\ref{Sdef}), in the two-charge case we find
\begin{equation}\label{Ssep2ch}
\Sep_{\rm 2ch}=G+2\Ared=X+2\Ared-a^2\sin^2\theta~.
\end{equation}
In the notation (\ref{Ssep}), we find $\Sep_{R{\rm 2ch}}=X+2\Ared$ and $\Sep_{\Theta{\rm 2ch}}=-a^2\sin^2\theta$. Thus the CKSTs given in (\ref{P4d}) and (\ref{S4d}) become:
\begin{align}
P^{\mu\nu}_{\rm 2ch} \partial_\mu\partial_\nu &= \partial_\theta^2 + \frac{1}{\sin^2\theta}\left( \partial_\phi + a\sin^2\theta\partial_t\right)^2
\label{P2ch}~,\\
S^{\mu\nu}_{\rm 2ch} \partial_\mu\partial_\nu  &= X\partial_r^2-\frac{1}{X}\left( (X+\Ared)\partial_t
+ a\partial_\phi\right)^2~. \label{S2ch}
\end{align}
Collecting formulae, we finally find the Exact Killing Stackel Tensor for the two-charge case:
\begin{align}
\Omega_{\rm 2ch} K^{\mu\nu} =& (X+\Ared-a^2)\left(\partial_\theta^2 + \frac{1}{\sin^2\theta}\left( \partial_\phi + a\sin^2\theta\partial_t\right)^2 \right) \notag\\
\label{K2cheqn}
&- a^2\cos^2\theta \left(X\partial_r^2-\frac{1}{X}\left( (X+\Ared)\partial_t
+ a\partial_\phi\right)^2\right).
\end{align}
This expression can be recast in many equivalent ways.  Note when comparing to other expressions for the EKST that addition of $g^{\mu\nu}$ or symmetric products of the Killing vectors will change the form of $K^{\mu\nu}$.

We also note that in the Kerr case, $\Ared$ becomes just $2mr$; our coordinates match the Boyer-Lindquist coordinates, and the EKST takes the standard form in those coordinates.

\subsubsection{The 5D mSUGRA Black Holes and the Dilute Gas Limit}
We next consider the case where three charges are equal:
\begin{equation}
\delta_1=\delta_2=\delta_3~.
\end{equation}
This special case can be embedded simply into supergravity, by reducing 5D minimal supergravity to 4D. This case is also related to the dilute gas black holes and the subtracted
black holes with modified asymptotic behavior.

Since a CKST exists for four independent charges it also exists in this special case. The natural question is thus whether there is any enhancement to an EKST. The criterion we consider is again whether the conformal
factor $\Omega=\sqrt{\Delta}$ separates as indicated in (\ref{Omegasep}). In the case of three equal charges the conformal factor $\Delta_0$ (\ref{Delta0rewritten}) can be simplified somewhat and written as
\begin{align}\label{Delta031}
\Delta_{0,3+1} &= (G+\Ared)^2-8 m \sinh\left(\frac{\delta_1-\delta_0}{2}\right)^3\sinh\left(\frac{1}{2}(3\delta_1+\delta_0)\right)r G\\
&+8 m^2 \left[\cosh\left(\frac{1}{2}(3\delta_1-\delta_0)\right)+3\cosh(\frac{\delta_1+\delta_0}{2})\right]\sinh\left(\frac{\delta_1-\delta_0}{2}\right)^3\sinh(\delta_1)^3G~.\notag
\end{align}
For $\delta_0=\delta_1$ the right hand side is indeed a complete square and, upon taking the square root, the separability property (\ref{Omegasep}) is satisfied. This is unsurprising, since this special case reduces to Kerr-Newman, a subclass of the two-charge case we studied in the previous subsection. Importantly, there is no corresponding simplification for generic values of $\delta_0$ and $\delta_1$: the right hand side of (\ref{Delta031}) is not generally a complete square and so $\Omega=\sqrt{\Delta}$ does not separate.

It is worth verifying this claim in more detail in the dilute gas limit, where the three charges with value $\delta_1$ are large. We explicitly implement this limit by taking
\begin{equation}
\delta_1\rightarrow \infty~, ~~\partial_t\rightarrow 0~, ~~e^{3\delta_1}\partial_t = {\rm fixed}~.
\label{DGlimit}
\end{equation}
The remaining parameters $m, a, r, \delta_0$ are also kept fixed in the dilute gas limit.

In this limit we find that the effective potential $\Sep$ reduces to a constant:
\begin{equation}
\Sep_{0,3+1}\partial_t^2\rightarrow \Sep_{0,DG}\partial_t^2=-\frac{m^2}{16}e^{6\delta_1-2\delta_0}\partial_t^2~.
\end{equation}
This simplification implies that the Laplacian on the right hand side of (\ref{4chinvmetric}) reduces to hypergeometric form, a remnant of the AdS$_3$ interpretation underlying this limit.
In this manner the wave equation has simplified considerably.

The question of whether the dilute gas limit allows an EKST is again determined by separability of $\Omega=\sqrt{\Delta}$, where now (\ref{Delta031}) becomes
\begin{align}
\Delta_{0,DG} &= \frac{1}{ 16} m^2 e^{6\delta_1} \left(  \left( r \cosh\delta_0 - (r-2m)\sinh\delta_0 \right)^2
- e^{-2\delta_0} G\right) \notag \\
&= \frac{1}{ 16} m^2 e^{6\delta_1} \left(  2mr + 4m^2 \sinh^2\delta_0 - a^2 e^{-2\delta_0}\cos^2\theta \right)~.
\label{DeltaDG}
\end{align}
The expression in the final bracket is evidently not a complete square. This in turn prevents separability of $\Delta^{1/2}$ and so precludes the construction of an EKST satisfying (\ref{KSeqn}).

\subsubsection{The Subtracted Black Holes}
As we have previously mentioned, we also wish to consider the existence of Killing tensors for the subtracted spacetimes. These geometries again take the form (\ref{metric}), but now with the conformal factor $\Delta_s$ given in (\ref{DeltaS}). The construction of CKSTs leading to $P^{\mu\nu}$ as in (\ref{P4d}) or, equivalently, to $S^{\mu\nu}$ as in (\ref{S4d}), is entirely unchanged. Moreover, the corresponding inhomogeneous terms are still given by  (\ref{Veqn}) and (\ref{Ueqn}), although these expressions should of course be computed from the conformal factor of the subtracted geometry, {\it i.e.}
$\Delta_s$ as given in (\ref{DeltaS}).

The subtracted geometries were constructed in \cite{Cvetic:2011dn}, by demanding that they promote certain approximate near horizon symmetries to exact symmetries. Despite this enhanced symmetry, the CKSTs of the subtracted geometries cannot be promoted to EKSTs. As in previous examples, this is
evident from the fact that the subtracted conformal factor (\ref{DeltaS}) fails to separate in the manner
demanded in (\ref{Omegasep}).

We do notice a relationship between the separated geometries and the dilute gas regime.  Both cases have constant effective potential $\Sep$, and consequently have the same functional form for the conformal factor $\Delta$, as we can see from (\ref{DeltaS}) and (\ref{DeltaDG}). This is no accident, as the subtracted geometries can be viewed as generalizations of the dilute gas geometries to the setting beyond the dilute gas regime.

As both the subtracted geometries and the dilute gas regime have a natural 5D lift, one could examine the nature of the lift of the CKSTs present in these cases.  We leave this exploration to future work.

\subsection{Quantum Separability}\label{quantumsection}
The strongest form for separability of the Klein-Gordon equation is {\it quantum separability}. This refers to the situation where the differential operator constructed from the EKST commutes with the d'Alembertian:
\begin{equation}\label{commutatorcond}
[\nabla^{\mu}K_{\mu\nu}\nabla^{\nu}~, ~g_{\lambda\sigma}\nabla^{\lambda}\nabla^{\sigma}]=0~.
\end{equation}
This condition is the strongest in the sense that it makes no distinction between massive and massless Klein-Gordon; and it obviously implies separability of the geodesic equations as well.

It is straightforward to expand the commutator (\ref{commutatorcond}) under the assumption that $K_{\mu\nu}$ is an EKST satisfying (\ref{KSeqn}). The commutators of covariant derivatives give rise to terms involving the Riemann tensor which do not automatically cancel. Instead they impose the condition
\cite{Carter:1979fe}
\begin{equation}\label{quantcond}
K_\nu^{[\rho}R^{\mu]\nu}=0~,
\end{equation}
on the EKST.

It was noted already by Carter that the quantum condition (\ref{quantcond}) is in fact satisfied for the Kerr-Newman black hole, where we set all four charges
equal. We find that it is also satisfied for the two-charge EKST constructed in (\ref{K2cheqn}).

\subsection{Summary of Killing-Stackel Results}
In this section, we have shown that the two-charge, Kerr-Newman, and Kerr black holes all possess a nontrivial EKST, in addition to a family of CKSTs. Consequently these backgrounds have a separable Hamilton-Jacobi equation for both the massless and massive cases,
as well as a separable massive Klein-Gordon equation.

In contrast, the full four-charge background only has a family of CKSTs. The 5D mSUGRA reduction black holes also do not have special properties beyond those of the four-charge background.  Similarly, both the dilute gas limit and the subtracted backgrounds possess the same Killing structure as the full four-charge background.

\section{Shearfree Null Congruences}\label{shearfreesection}
It has been known since their construction \cite{Newman:1965my} that Kerr-Newman black holes possess a pair of null congruences that are both geodesic and shearfree. These congruences are related in a natural way to the Conformal Killing-Stackel tensors.
In this section we show that these properties generalize to two-charge black holes but not to generic four-charge black holes.

\subsection{Null Congruences for Two-charge metrics}\label{shearfree2charge}
The separation (\ref{uppermetricseparable}) of the inverse metric into CKSTs $P^{\mu\nu}(\theta)$ and $S^{\mu\nu}(r)$ amounts to a block diagonalization of the metric $g_{\mu\nu}$ into a two dimensional angular space with Euclidean metric $P_{\mu\nu}$ and a two dimensional radial/temporal space with Lorentzian metric $S_{\mu\nu}$.

Apart from the overall factor $\Omega$ in (\ref{uppermetricseparable}) the decomposition is a projection onto 2D surfaces.  In the two-charge case
the radial/temporal space is naturally decomposed as
\begin{equation}
S^{\mu\nu} = -2\Omega l^{(\mu} n^{\nu)}~,\label{Sintoln}\\
\end{equation}
where $l^\mu$ and $n^\nu$ are real null vectors. It is natural to expand the angular space analogously, as
\begin{equation}
P^{\mu\nu} = 2\Omega m^{(\mu}\bar{m}^{\nu)}~.\label{Pintommbar}
\end{equation}
Since the angular space is Euclidean, the ``lightcone'' basis vectors $m^\mu$ and $\bar{m}^\mu$ must be complex. Indeed,
they are complex conjugates of each other.

The vectors $l^\mu$ and $n^\nu$ are null in the radial/temporal space and they are orthogonal to the angular basis vectors $m^\mu, \bar{m}^\mu$; so they are null in the full geometry. The angular basis vectors are similarly null in the full geometry so, taken together, the four vectors $l^\mu,n^\mu,m^\mu, \bar{m}^\mu$ in fact produce a complex null tetrad that factorizes the metric as
\begin{equation}\label{metrictetrad}
g^{\mu\nu} = -2l^{(\mu}n^{\nu)}+2m^{(\mu}\bar{m}^{\nu)}~.
\end{equation}
It is clear from this form the null tetrad has been normalized canonically
\begin{equation}\label{nulltetrad}
m\cdot\bar{m}=1=-l\cdot n~,
\end{equation}
in addition to the null conditions $l^2=n^2=m^2=\bar{m}^2=0$ and
the orthogonality conditions $m\cdot l= m\cdot n = \bar{m}\cdot l=\bar{m} \cdot n =0$.

The radial null vector $l_\mu=\partial_\mu S$ can be interpreted as a generator of solutions to the massless HJ equation. As such it is equivalent to a geodesic. There is one such geodesic for each point in the angular space so these geodesics cover the entire spacetime, ie. they form a congruence. The $l^\mu$ congruence of ingoing geodesics is complemented by a $n^\nu$ congruence of outgoing geodesics.

Explicitly, in the two-charge backgrounds, the radial CKST $S^{\mu\nu}$ is represented in a form (\ref{S2ch}) that can be readily recast as
a symmetrized product of two real null vectors as in (\ref{Sintoln}), where
\begin{align}
l^\mu & = \frac{1}{X}\left(\Ared+X,X,0,a\right)~,\label{l2ch}\\
n^\mu & = \frac{1}{2\Omega}\left(\Ared+X,-X,0,a\right)~.\label{n2ch}
\end{align}
Indices are written in the order $(t,r,\theta,\phi)$.
The angular CKST $P^{\mu\nu}$ (\ref{P2ch})  is represented in a form that can be similarly recast as a symmetrized product of two complex null vectors
\begin{align}
m^\mu & = \frac{1}{\sqrt{2\Omega}}(a\sin\theta,0,i,\frac{1}{\sin\theta})~,\label{m2ch}\\
\overline{m}^\mu & = \frac{1}{\sqrt{2\Omega}}(a\sin\theta,0,-i,\frac{1}{\sin\theta})~.\label{mbar2ch}
\end{align}

The null tetrad $(n,l,m,\bar{m})$ decomposes the CKSTs as (\ref{Sintoln},\ref{Pintommbar}). The EKST (\ref{K2cheqn}) for the two-charge case
can then be expressed economically as:
\begin{align}\label{killingtetrad}
K^{\mu\nu} &= -2f(\theta)l^{(\mu}n^{\nu)}-2 h(r)m^{(\mu}\bar{m}^{\nu)}\\
&= -2a^2\cos^2\theta l^{(\mu}n^{\nu)}-2 \left(X+\Ared-a^2\right)m^{(\mu}\bar{m}^{\nu)}~.  \notag
\end{align}

As a null congruence evolves its rays move relative to each other in the two dimensional plane that is orthogonal to the congruence without being along the null ray. This two dimensional plane is precisely the angular space spanned by $m^\mu$, $\bar{m}^\mu$. The shear of the null congruence is the symmetric traceless part of the $2\times 2$ matrix characterizing the evolution, so the shear is captured by projection onto $m^\mu m^\nu$ and its complex conjugate. By direct computation we find that the shear in fact vanishes
\begin{align}
\sigma &= -m^\mu m^\nu \nabla_\mu n_\nu = 0~, \label{sigmaeqn}\\
\lambda &= \bar{m}^\mu \bar{m}^\nu \nabla_\mu l_\nu = 0~. \label{lambdaeqn}
\end{align}
In these formulae $\sigma, \lambda$ is the conventional Newman-Penrose notation for the spin coefficients encoding shear of the congruences tangent to $n^{\mu}, l^{\mu}$ respectively.

Shear is an interesting quantity to study because of its prominent role in the study of exact solutions, especially algebraically special solutions.  Specifically, the well-known Goldberg-Sachs theorem relates the presence of a shearfree, geodesic null congruence in a vacuum spacetime to algebraic speciality.  As we will explore more thoroughly below, non-empty spacetimes do not have such a direct relationship; the Mariot-Robinson theorem shows is possible to have a shearfree geodesic null congruence in a non-empty spacetime which is not algebraically special, provided the matter present satisfies particular conditions.

Also, as we address in Section \ref{KYsection}, nondegenerate Killing-Yano tensors are always constructed from shearfree geodesic congruences.  Thus, if we would like to discuss these objects, we should first try to construct shearfree congruences.

As we have already shown, the vectors $n$ and $l$ are tangent to geodesics; thus, given their shearfreeness they in fact are tangent to geodesic shearfree null congruences.  We can encode their geodesicness in a different set of spin coefficients:
\begin{align}
\kappa &= -m^\nu n^\mu \nabla_\mu n_\nu = 0~. \label{kappaeqn}\\
\nu &= \bar{m}^\nu l^\mu \nabla_\mu l_\nu = 0~, \label{nueqn}
\end{align}
Thus 4 (out of 12) Newman-Penrose spin coefficients vanish for the natural null congruences in the two-charge black hole backgrounds.

In the above, we introduced the two natural null congruences as the eigenvectors of the CKSTs and then showed that they were shearfree by direct computation. An alternative route that brings both properties at one go is to apply the Mariot-Robinson theorem \cite{Mariot,Robinson} (Theorem 7.4 in \cite{Stephani:2003tm}) which shows the existence of a shearfree null congruence for every distinct field strength $F^{\mu\nu}$ which solves the Maxwell equations and is also null, defined for two forms as the property   $F_{\mu\nu}F^{\mu\nu}=F^*_{\mu\nu}F^{\mu\nu}=0$. For the two-charge black holes an appropriate field strength is
\begin{align}
F_{tr} &= \frac{a}{X}~,\; F_{r\phi} = 1+\frac{\Ared}{X}~, \; F_{\phi t} = 1~, \notag
\\
F^*_{\theta t} &=\frac{1}{\sin\theta}~, \; F^*_{r\theta} = \frac{\sqrt{\Delta_{2ch}}}{X\sin\theta}~, \; F^*_{\phi\theta} = a\sin\theta~. \label{F2ch}
\end{align}
The corresponding null congruence identified through the zero-mode conditions
\begin{equation}
F_{\mu\nu}l^{\nu} = 0~, \; F^*_{\mu\nu}l^\nu =0~,\notag
\end{equation}
is the same as $l^\mu$ already given in (\ref{l2ch}). The Mariot-Robinson theorem then guarantees that this null congruence is shear-free. The analogous ``test'' tensor for the outgoing null congruence $n^{\mu}$ is determined from (\ref{F2ch}), by changing the sign of all tensor components that include an $r$ index.

\subsection{Null Congruences for Generic Black Holes}
The simplifications enjoyed by the null congruences of two-charge black holes only apply in part to the more general black holes described by metrics of the form (\ref{metric}).

The general black holes do of course have null geodesics and we can find them by exploiting separability (\ref{uppermetricseparable}).
The Killing vectors ensure the constancy of the energy $E=l_t$ and the angular momentum $J=l_\phi$. The Hamilton-Jacobi equation then determines the remaining components of $l_\mu$ as
\begin{align}
l_r & = \sqrt{ \frac{1}{X^2}\left( \Ared E + aJ\right)^2 + \frac{1}{X}(\Sep_R+\Sep_c)E^2- \frac{1}{X}J^2}~,\label{genlr}\\
l_\theta & = \sqrt{(\Sep_\Theta  - \Sep_c)E^2- \frac{\cos^2\theta}{\sin^2\theta} J^2}~.\label{genltheta}
\end{align}
The separation constant $\Sep_c$ and the conserved quantities $E, J$ provide three independent integration constants, as
there should be for general null geodesics. Written in this form, (\ref{genlr}) and (\ref{genltheta}) provide the full set of ingoing null geodesics within any separable background of the form (\ref{metric}).  In the following, we will focus on the asymptotically flat four-charge backgrounds, but a similar procedure with the same result may be applied to more general cases, such as the subtracted backgrounds or the Kaluza-Klein black holes.

The HJ equation implies the geodesic equation $l^\mu\nabla_\mu l_\nu=0$ but we can
also check the geodesic equation explicitly: it reduces to the conditions $\partial_a l_i=0$ and $\partial_b l_a - \partial_a l_b=0$ where $i=t,\phi$ and $a=r,\theta$. These conditions further ensure that the geodesics are hypersurface forming, equivalent to the vanishing of their rotation $\widehat{\omega}_{\mu\nu}=0$.

Among the full set of null geodesics, we are particularly interested in those which are orthogonal to the CKST $P^{\mu\nu}$, because we wish to make decompositions as in (\ref{Sintoln},\ref{Pintommbar}). In fact the four-charge $P^{\mu\nu}$ given in (\ref{P4d}) with $\Sep_\Theta=-a^2\sin^2\theta$ does not depend on charges at all. We can therefore use the two-charge
result (\ref{m2ch},\ref{mbar2ch}) again and will then have decomposed $P^{\mu\nu}$ into $m, \bar{m}$ just as in (\ref{Pintommbar}) for the two-charge case.  However, here we hit our first
snag: the complex vectors $m,\bar{m}$ given in (\ref{m2ch},\ref{mbar2ch}) have norm
\begin{equation}
2\Delta m\cdot m = 2\Delta \bar{m}\cdot\bar{m} = \frac{a^2\sin^2\theta}{G} \left( \Delta - (G+\Ared)^2\right)~.
\end{equation}
In the two-charge case the expression vanishes as it should, because then $\Delta_{\rm 2ch}=(\Ared  + G)^2$. However, as we already detailed in (\ref{Delta0rewritten}), this identity is special to the two-charge case. Thus the proposed angular vectors $m,\bar{m}$ are not null in the general four-charge case.

Proceeding nonetheless, we seek a radial vector $l$ that is orthogonal to $P^{\mu\nu}$. In view of the decomposition (\ref{Pintommbar}) it is sufficient to demand orthogonality with $m,\bar{m}$. This condition gives $l_\theta=l_\phi+a\sin^2\theta l_t=0$ which gives  $l^\theta=0$, ie. each
null ray evolves along a fixed value of $\theta$. Using (\ref{genltheta}) to determine the separation constant $\Sep_c$, $l_\theta=J+a\sin^2\theta E=0$ to determine $J$, and using scaling symmetry to take $E=-1$, we can recast (\ref{genlr}) as
\begin{align}
l_r & = \frac{\sqrt{\Delta}}{ X} \sqrt{1+ \frac{a^2\sin^2\theta}{\Delta G} \left(\Delta - (\Ared  + G)^2\right)}~.\label{lr4ch}
\end{align}
The resulting vector $l_\mu=(-1,l_r,0,a\sin^2\theta)$ with $l_r$ given in (\ref{lr4ch}) is a null eigenvector for the four-charge CKST $P^{\mu\nu}$, by construction; but it is also a null eigenvector for the four-charge CKST $S^{\mu\nu}$ given in (\ref{S4d}), as we hoped for.

The expression (\ref{lr4ch}) for $l_r$ is quite complicated algebraically but it simplifies greatly in the two-charge case where
we can use $\Delta_{\rm 2ch}=(\Ared  + G)^2$. This apparent complication of the four-charge case is even more striking after using
the metric (\ref{metric}) to determine the covariant form of the radial vector
\begin{align}
l^\mu  &= \left(
\frac{G+\Ared }{\sqrt{\Delta}} ( \frac{\Ared }{ X} + 1)+ \frac{\Delta - (G+\Ared  )^2}{ G\sqrt{\Delta}},~
\sqrt{ 1+ \frac{a^2\sin^2\theta}{\Delta G}\left(\Delta - (G+\Ared )^2\right)}, \right. \notag \\
& ~~~~~
\left. 0,~ \frac{a}{ X}\frac{G+\Ared }{\sqrt{\Delta}}
\right)~.
\end{align}
In the general four-charge case separability implies that $\Delta-(\Ared  + G)^2$ can be factorized by $G$, but this property does not give simplifications in the formulae above.

In order to attempt to fully diagonalize the CKST $S^{\mu\nu}$, we should also give the outgoing congruence $n^\mu$. It takes the form
$n_\mu \propto (-1,-l_r,0,a\sin^2\theta)$ with $l_r$ still given by (\ref{lr4ch}) and the overall normalization determined such that
$l\cdot n=-1$. The normalization condition turns out to involve the awkward combination appearing under the square root in (\ref{lr4ch})
such that the final expression for the properly normalized ingoing null congruence becomes
\begin{align}
n^\mu  &= \frac{1}{ 2\left[\Delta+ \frac{a^2\sin^2\theta}{ G}\left(\Delta - (G+\Ared )^2\right)\right]}\left(
\left(G+\Ared \right) ( \Ared + X)+ X \frac{\Delta - (G+\Ared  )^2}{ G},\right. \notag \\
& ~~~~~\left.
-X\sqrt{ \Delta+ \frac{a^2\sin^2\theta}{G}\left(\Delta - (G+\Ared )^2\right)},~
0,~ a\left(G+\Ared \right)
\right)~.
\end{align}

In summary, we have constructed a pair of hypersurface forming null congruences $l$ and $n$ that are orthogonal to the CKST $P^{\mu\nu}$ in the general four-charge case. As we have emphasized, they are much more complicated than in the two-charge case. In addition to this aesthetic and practical criterion we stress that they also suffer from several physically undesirable features:
\begin{itemize}
\item
As we have already noted, the angular vectors $m, \bar{m}$ are no longer null. Therefore $l, n, m, \bar{m}$ do not form a null tetrad.
\item
The radial CKST $S^{\mu\nu}$ cannot be expressed as the symmetric product of $n,l$ as we did for the two-charge case in (\ref{Sintoln}).
\item
The interpretation of the CKSTs $P^{\mu\nu}$ and $S^{\mu\nu}$ as projectors onto the angular and the radial space is undermined by the failure of these
tensors to be mutually orthogonal
\begin{equation}
S^{i \mu}P_\mu^{~j} = \frac{a^2\sin^2\theta}{\sqrt{\Delta}} \frac{(G+\Ared)^2 - \Delta}{ G} ~P^{ij}~,
\end{equation}
where $i,j$ run over $t,\phi$ as before.  Importantly, alternate choices of separation constants in (\ref{uppermetricseparable}) cannot remedy this deficiency.
\item
Both $l$ and $n$ have nonzero shear in the full four-charge background.
\end{itemize}

\section{Petrov Type D and Killing Yano Tensors}\label{KYsection}
Thus far we have not shown that the Kerr-Newman case produces any simplifications over the general two-charge case, nor that it possesses any Killing structure beyond that of the two-charge case. In this section we will show that the Kerr-Newman spacetime
is ``more special'' than the two-charge case, in at least two aspects:
\begin{itemize}
\item
It is {\it algebraically special}: the spacetime becomes Petrov Type D~.
\item
It allows a {\it Killing-Yano tensor}.
\end{itemize}
For the general two-charge case, the spacetime is not algebraically special and there is no Killing-Yano tensor.

\subsection{The Killing-Yano Tensor}
A Killing-Yano tensor is an {\it antisymmetric} two tensor $f_{\mu\nu}$ with square that equals an Exact Killing-Stackel Tensor $K_{\mu\nu}$:
\begin{equation}
K_{\mu\nu} = f_{\mu\rho}f^{\rho}_{\phantom{\rho}\nu} \label{KYtoKS}~.
\end{equation}
The Killing-Yano tensor must further satisfy the Killing-Yano equation:
\begin{equation}
\nabla_\mu f_{\nu\rho} + \nabla_{\nu} f_{\mu\rho}=0~. \label{KYeqn}
\end{equation}
In the event that the Killing-Yano equation (\ref{KYeqn}) has been solved, the square (\ref{KYtoKS}) will in fact satisfy the Killing tensor equation (\ref{KSeqn}). However, it is not always true that a Killing tensor can be written as the square of a Killing-Yano tensor.

Since Killing-Yano implies Killing-Stackel, we know from the outset that we may find a Killing-Yano only in the cases where an EKST exists. We therefore focus on the two-charge black holes for which we exhibited an EKST in (\ref{K2cheqn}). The best starting point is the expression (\ref{killingtetrad}) of the EKST in terms of the null tetrad because it provides a simple and systematic way to construct a ``square-root'' tensor: take the {\it antisymmetric} products of the tetrad basis and the usual square root of the coefficients in this basis. Proceeding this way we find
\begin{align}
f^{\mu\nu} &= -2\sqrt{f(\theta)}l^{[\mu}n^{\nu]}+2i\sqrt{h(r)}m^{[\mu}\bar{m}^{\nu]}\label{KYeasy}\\
&= -2a\cos\theta l^{[\mu}n^{\nu]}+2i\sqrt{X+\Ared-a^2}m^{[\mu}\bar{m}^{\nu]}~, \notag
\end{align}
where $f^{\mu\nu}$ is the anti-symmetric two-tensor which has a square (\ref{KYtoKS}) that produces the EKST in (\ref{K2cheqn}).
This object takes a particularly simple form when it is expressed with lower indices, {\it i.e.} as the two-form
\begin{align}\label{f2form}
f =&~ a \cos \theta~ dr \wedge\left(dt-a\sin^2\theta d\phi\right)\\
&-\sin\theta\sqrt{X+\Ared-a^2}\left(-a dt+(X+\Ared)~d\phi\right)\wedge d\theta~.\notag
\end{align}
In particular for Kerr black holes where all charges are tuned to zero we recover the Killing-Yano tensor in Boyer-Lindquist
coordinates (e.g. equation (9) of \cite{Bredberg:2011hp}).

The $f_{\mu\nu}$ exhibited here is the ``square-root'' of the EKST, by construction; but it does {\it not} necessarily satisfy the Killing-Yano equation (\ref{KYeqn}). We find that instead the tensor $\nabla_\mu f_{\nu\rho}+\nabla_\nu f_{\mu\rho}$ becomes proportional to $\sqrt{X+\Ared-a^2}-\partial_r\Ared-\partial_r X$; for example, we find
\begin{equation}
\nabla_t f_{\theta\theta}+\nabla_\theta f_{t\theta}=\nabla_\theta
f_{t\theta}=\frac{a \cos\theta X}{2 \sqrt{\Delta}}\left(\sqrt{X+\Ared-a^2}-\partial_r\Ared-\partial_r X\right)~.
\end{equation}
In the two-charge case we have the simple expression
\begin{equation}
X+\Ared-a^2 = (r + 2m\sinh^2\delta_1)(r + 2m\sinh^2\delta_2)~,\label{xared}
\end{equation}
which allows us to write
\begin{equation}\label{residual}
\nabla_\mu f_{\nu\rho}+\nabla_\nu f_{\mu\rho} \sim \sqrt{(r + 2m\sinh^2\delta_1)(r + 2m\sinh^2\delta_2)} -r-m\sinh^2\delta_1-m\sinh^2\delta_2~.
\end{equation}

For general two-charge black holes, this expression is manifestly nonzero. Only in the special case of Kerr-Newman black holes, {\it i.e.} when the two charges are equal, does this expression vanish. This is the special case where the combination (\ref{xared}) is a complete square and only in this situation does the candidate $f_{\mu\nu}$ (\ref{f2form}) become a true Killing-Yano tensor.

We can confirm our conclusion that a true Killing-Yano tensor exists only in the Kerr Newman case via an indirect argument involving the Petrov type. As shown in \cite{Taxiarchis} (and also Chapter 35 in \cite{Stephani:2003tm}), a spacetime may only have a true Killing-Yano tensor if it is algebraically special. In fact, the spacetime must be at least of Petrov type D. As we will show in the following section, this condition is so strong that, among generic two-charge black holes, it is satisfied only in the Kerr-Newman case. This conforms with our explicit finding that the candidate Killing-Yano tensor (\ref{f2form}) is an actual Killing-Yano tensor only in the Kerr-Newman case.

Even though the lack of algebraic speciality for the general two-charge case prevents us from constructing a true Killing-Yano tensor, the separability of the massless wave equation and the associated EKST does lead us to a unique candidate Killing-Yano tensor even in the two-charge case. This motivates the hope that some generalized Killing-Yano equation exists which is useful more broadly. The concept of conformal Killing-Yano tensor is not useful, because it also requires spacetimes to be algebraically special \cite{Glass:1998ka, Mason}. However, other notions of generalized Killing-Yano tensors exist, such as those introduced in \cite{Frolov:2008jr,Kubiznak:2009sm,Kubiznak:2009qi,Wu:2009cn,Houri:2012eq}, and such as the obvious ``square-root'' of the CKSTs constructed in this paper. It would be interesting to determine if such notions are useful for
various classes of black holes in string theory.

\subsection{Petrov Classification}
The starting point for Petrov's algebraic classification of spacetimes is the Weyl tensor $C$, the traceless part of the Riemann curvature tensor. In four dimensions the Weyl tensor allows up to four principal null vectors, defined as solutions $k^\mu$ of the generalized eigenvalue equation
\begin{equation}\label{principalnullvectoreqn}
k_{[\mu}C_{\nu]\rho\sigma[\tau}k_{\chi]}k^\rho k^\sigma=0~,~~ k^\mu k_\mu=0~.
\end{equation}
Algebraic speciality, or higher Petrov type, occurs when two or more principal null vectors coincide. Petrov type D is the case where the four principal null vectors are identical in pairs.
One way to simplify the condition (\ref{principalnullvectoreqn}) for practical computations is to take advantage of the null tetrad which gives the equivalent condition
\begin{equation}\label{Cprincipaleqn}
C_{\mu\nu\rho\sigma}k^\mu m^\nu k^\rho m^\sigma=0~,
\end{equation}
where $m^\mu$ is a complex null vector orthogonal to $k^\mu$. For our purposes it will be sufficient to simplify further and exploit an indirect method. Following the
Newman-Penrose formalism, the Weyl tensor amounts to the five complex components
\begin{align}
\Psi_0 &= C_{\mu\nu\rho\sigma} n^\mu m^\nu n^\rho m^\sigma~, \notag\\
\Psi_1 &= C_{\mu\nu\rho\sigma} n^\mu l^\nu n^\rho m^\sigma~, \notag\\
\Psi_2 &= -C_{\mu\nu\rho\sigma} n^\mu m^\nu l^\rho \bar{m}^\sigma~, \label{Psidefs}\\
\Psi_3 &= C_{\mu\nu\rho\sigma} l^\mu n^\nu l^\rho \bar{m}^\sigma~, \notag\\
\Psi_4 &= C_{\mu\nu\rho\sigma} l^\mu \bar{m}^\nu l^\rho \bar{m}^\sigma~, \notag
\end{align}
where $n,~l,~m,~\bar{m}$ refer to the null tetrad. The indirect method relies on each of the principal null vectors $k$ being associated with a complex root $E$ of the eigenvalue equation
\begin{equation}\label{eigenvalue}
\Psi_0-4 \Psi_1 E+ 6 \Psi_2 E^2-4 \Psi_3 E^3+\Psi_4 E^4=0~.
\end{equation}
If any of these roots coincide, so do their associated principal null vectors. Thus we can leverage knowledge of $\Psi_i$ to understand whether a spacetime is algebraically special.

The Newman-Penrose equations express (among other things) the Weyl tensor components as differential operators acting on spin-coefficients. Specifically, the Weyl invariant $\Psi_0$ depends linearly on the spin coefficients $\sigma$ and $\kappa$; and $\Psi_4$ depends linearly on the spin coefficients $\nu$ and $\lambda$. These four spin coefficients all vanish, according to (\ref{sigmaeqn}-\ref{nueqn}), so the Newman-Penrose equations show that $\Psi_0=\Psi_4=0$. Comparing the definitions (\ref{Psidefs}) with the criterion (\ref{Cprincipaleqn}), we see that $n^\mu$ is a principal null vector because $\Psi_0$ vanishes and $l^\mu$ is a principal null vector because $\Psi_4$ vanishes.

We can also compute $\Psi_1$ and $\Psi_3$, finding:
\begin{align}
\Psi_3 &= \frac{4am^2 (\sinh^2\delta_1-\sinh^2\delta_2)^2\sin\theta}{(2\Omega)^{5/2}}~, \notag\\
\Psi_1 &= \frac{X}{2\Omega}\Psi_3~.
\end{align}
Insofar as these expressions for $\Psi_1$ and $\Psi_3$ take generic finite values, (\ref{eigenvalue}) will then have four distinct roots: $0$, $\infty$, and the two finite roots of a quadratic equation. Thus generic two-charge black holes do not have any repeated principal null vectors and so they are not algebraically special; they only Petrov type I, the generic type. Note that for the full two-charge case we have not actually found the principal null vectors beyond $n$, $l$. However, we have deduced their existence from the nonzero $\Psi$ components.

The only special case that evades this generic conclusion is when $\Psi_3=\Psi_1=0$, a circumstance the arises if and only if $\delta_1=\delta_2$, ie. the remaining two charges are equal, corresponding to the Kerr-Newman case. Then the two finite roots of the generic quadratic equation degenerate to $0$ and $\infty$. In other words, the quartic equation (\ref{eigenvalue}) has two double roots. We can see this directly as well, by noting that when all charges are equal only $\Psi_2$ is nonvanishing, so $0$ and $\infty$ are double roots. Either way, we conclude that the principal null vectors $n^\mu$ and $l^\nu$ both become repeated principal null vectors in the Kerr-Newman case.  By definition, the spacetime is then Petrov type D.

%

%
%
\section{Separability of Kaluza-Klein Black Holes}\label{magchargesection}
Thus far, we have considered only the set of four-charge black holes found in \cite{Cvetic:1995kv}, and their subtracted versions \cite{Cvetic:2011dn}.  However these four-charge black holes are not the most general possible rotating black holes in $N=4,8$ string theory \cite{Cvetic:1996zq}. Although there is not an explicit construction available for the wider class of black holes, the solution generation technique provides an indirect specification of these geometries.

In this section, we will explore the possibility that this solution generation mechanism preserves the separability structure.  Specifically, we study the Killing structure and separability present in the rotating Kaluza-Klein black hole constructed in \cite{Larsen:1999pp}
(see also \cite{Horowitz:2011cq}). It is parameterized by the mass $M$, the angular momentum $J$, and the electric (magnetic) charges $Q(P)$ under the Kaluza-Klein vector field. These physical parameters are in turn expressed in terms of parametric mass $m$, parametric angular momentum $a$, and parametric charges $q,p$ through
\begin{align}
2G_4 M & = \frac{p+q}{ 2} ~,\\
G_4 J & = \frac{\sqrt{pq} (pq + 4m^2)}{ 4m (p+q)} a~,\\
Q^2 & = \frac{q(q^2-4m^2)}{ 4 (p+q)}~,\\
P^2 & = \frac{p(p^2-4m^2)}{ 4 (p+q)}~.
\end{align}

The metric of the Kaluza-Klein black holes can be presented as \cite{Larsen:1999pp}\footnote{The notation used here is an obvious redefinition of the one used in \cite{Larsen:1999pp}:
$G_{\rm here} = H_{3, {\rm there}}$, $X_{\rm here} = \Delta_{\rm there}$, $A_{\phi,{\rm here}}=B_{\phi,{\rm there}}$.
It is worth stressing that the conformal factor $\Delta$ ubiquitous in the present work and defined for Kaluza-Klein black holes in (\ref{Finntohere}) is unrelated to the $\Delta$ used in \cite{Larsen:1999pp}.}:
\begin{equation}\label{KKmetric}
ds_4^2=-\frac{G}{\Delta^{1/2}}\left(dt+\frac{a \sin^2\theta}{G} \Ared d\phi\right)^2 +\Delta^{1/2}\left(\frac{dr^2}{X}+d\theta^2+\frac{X}{G}\sin^2\theta d\phi^2\right)~,
\end{equation}
which is the same form as (\ref{metric}) used earlier for the four-charge black holes. The functions
$G$ and $X$ remain unchanged from the earlier case (\ref{metricdefs}). The reduced potential for rotation is again
a function of radius alone
\begin{align}
\Ared &= \frac{G}{a\sin^2\theta} A_\phi= \sqrt{pq}\frac{\left(pq+4m^2\right)r-m(p-2m)(q-2m)}{2m(p+q)}~,\label{Breddef}
\end{align}
and the conformal factor is
\begin{align}
\Delta &= H_1 H_2~,\label{Finntohere}
\end{align}
where
\begin{align}
H_1 &=r^2+a^2\cos^2\theta+r(p-2m)+\frac{p}{p+q}\frac{(p-2m)(q-2m)}{2} \notag\\
&-\frac{p}{2m(p+q)}\sqrt{\left(q^2-4 m^2\right)\left(p^2-4m^2\right)}a \cos\theta~,\label{H1def}\\
H_2 &=r^2+a^2\cos^2\theta+r(q-2m)+\frac{q}{p+q}\frac{(p-2m)(q-2m)}{2}\notag \\
&+\frac{q}{2m(p+q)}\sqrt{\left(q^2-4 m^2\right)\left(p^2-4m^2\right)}a \cos\theta~.\label{H2def}
\end{align}

Since the metric (\ref{KKmetric}) is identical to the four-charge metric (\ref{metric}) with $G, X$ unchanged, the inverse metric remains as in (\ref{4chinvmetric}). As before, the only term that might obstruct separability is the effective potential
\begin{equation}\label{Bsep}
\Sep_{\rm KK}=\frac{\Delta - \Ared^2}{G}=\Sep_{R, {\rm KK}}(r) + \Sep_{\Theta, {\rm KK}}(\theta)~,
\end{equation}
where
\begin{align}\label{VeffRKK}
\Sep_{R, \rm KK}(r) &=-\frac{1}{4m^2(p+q)^2}\left[p^3q^3-8m^4\left(p^2+q^2\right)+4m^3(p+q)^2(p+q+2r)\right.\\
&\left.-2m^2\left(2q^2 r (q+r)+p^3(3q+2r)+2p^2(q^2+3qr+r^2)+pq(3q^2+6qr+4r^2)\right)\right]~.\notag
\end{align}
and
\begin{align}\label{VeffTKK}
\Sep_{\Theta, {\rm KK}}(\theta)=a\cos\theta\frac{(p-q)}{2m(p+q)}\sqrt{\left(p^2-4m^2\right)\left(q^2-4m^2\right)}+a^2\cos^2\theta~.
\end{align}
The important point is that the effective potential (\ref{Bsep}) separates into terms that depend on only $r$ and only $\theta$, respectively. Thus we recover a situation similar to the generic four-charge black hole:
\begin{itemize}
\item
The Kaluza-Klein black holes possess Killing vectors $\partial_t$ and $\partial_\phi$ which commute and exhibit hypersurface orthogonality, just as in Section \ref{Kvectorsection}.
\item
Separability of the potential (\ref{Bsep}) is tied to the existence of Conformal Killing Stackel Tensors given
by (\ref{P4d}) and (\ref{S4d}) with $\Sep_R$ and $\Sep_\Theta$ given in (\ref{VeffRKK}), (\ref{VeffTKK}).
\item
There is no Exact KST because $\Delta^{1/2}$ with $\Delta$ given in (\ref{Finntohere}) does not separate in this case (except when $H_1=H_2$ because $p=q$, and this is the Kerr-Newman geometry we have already considered in detail).
\end{itemize}
Thus the metric (\ref{KKmetric}) has separability properties most similar to those the generic four-charge background as in (\ref{Delta0}).
%
%
%
%

\subsection*{Acknowledgments}
We thank G. Gibbons for discussions.
This work was supported in part by US Department of Energy.

\newpage

\end{document}